\documentclass[12pt]{article}
\newwrite\ffile\global\newcount\figno \global\figno=1

\def\writedef#1{}

\input epsf
\def\figin{\epsfcheck\figin}\def\figins{\epsfcheck\figins}
\def\epsfcheck{\ifx\epsfbox\UnDeFiNeD
\message{(NO epsf.tex, FIGURES WILL BE IGNORED)}
\gdef\figin##1{\vskip2in}\gdef\figins##1{\hskip.5in}% blank space instead
\else\message{(FIGURES WILL BE INCLUDED)}%
\gdef\figin##1{##1}\gdef\figins##1{##1}\fi}

\def\figinsert{}
\def\ifig#1#2#3{\xdef#1{fig.~\the\figno}
\writedef{#1\leftbracket fig.\noexpand~\the\figno}%
\figinsert\figin{\centerline{#3}}\medskip\centerline{\vbox{\baselineskip12pt
\advance\hsize by -1truein\center\footnotesize{  Fig.~\the\figno.} #2}}
\bigskip\endinsert\global\advance\figno by1}
\def\endinsert{}

\begin{document}
\baselineskip 18pt
\newcommand{\Tr}{\mbox{Tr\,}}
\newcommand{\beq}{\begin{equation}}
\newcommand{\eeq}[1]{\end{equation}}
\newcommand{\bea}{\begin{eqnarray}}
\newcommand{\eea}[1]{\label{#1}\end{eqnarray}}
\renewcommand{\Re}{\mbox{Re}\,}
\renewcommand{\Im}{\mbox{Im}\,}
\begin{titlepage}

\begin{picture}(0,0)(0,0)
\put(350,0){SHEP-01-20}
\put(350,-15){}
\end{picture}

\begin{center}
\hfill
\vskip .4in
{\large\bf Superfluidity in the $AdS$/CFT Correspondence}
\end{center}
\vskip .4in
\begin{center}
{\large Nick Evans$^{a}$ and  Michela Petrini$^b$}
\footnotetext{e-mail: n.evans@hep.phys.soton.ac.uk, michela.petrini@unine.ch }
\vskip .1in
(a){\em Department of Physics, Southampton University, Southampton,
S017 1BJ, UK}
\vskip .1in
(b){\em Institute de Physique, Universite de Neuchatel, Neuchatel, Switzerland}
\end{center}
\vskip .4in
\begin{center} {\bf ABSTRACT} \end{center}
\begin{quotation}
\noindent A chemical potential may be introduced into the
$AdS$/CFT correspondence by setting the D3 branes of the construction spinning.
In the field theory the fermionic modes are expected to condense
as Cooper pairs, although at zero temperature the chemical
potential destabilizes the scalar sector of the ${\cal N}=4$
theory obscuring this phenomena. We show, in the case where a
chemical potential is introduced for a small number of the gauge
colours, that there is a metastable vacuum for the scalar fields
where fermionic Cooper pairing is apparently manifest.
In this vacuum the D3 branes expand non-commutatively (to balance the
centrifugal force) into a D5 brane, in a mechanism analogous to
Harmark and Savvidy's (M)atrix theory construction of a spinning
D2 brane. We show that the D5 brane acts as a source for the RR
3-form whose UV scaling and symmetries are those of a fermion
bilinear. The D5 brane rotates within the S$^5$ and so decays by
the emission of RR fields which we interpret as the metastable
vacuum decaying via higher dimension operators.
\end{quotation}
\vfill
\end{titlepage}
\eject
\noindent
\section{Introduction}

The $AdS$/CFT correspondence \cite{conjecture}, which is a duality
between a four dimensional strongly coupled gauge theory and a
weakly coupled gravity background (anti-de-Sitter space) in one
higher dimension, provides a new, controlled, framework in which
to describe the properties of strongly interacting systems. The
original duality for the conformal ${\cal N}=4$ super Yang-Mills
theory at the origin of moduli space resulted from consideration
of the limit where the gauge theory on the surface of $N$
coincident D3 branes decoupled from the supergravity  description
of the bulk spacetime. Many aspects of strong dynamics have since
been explored by deforming the ${\cal N}=4$ theory by the
inclusion of finite temperature \cite{w2} or relevant operators
\cite{5ddeform, ps, 10ddeform}. Motion in the radial direction of
the $AdS$ spacetime is interpreted as renormalization group flow
in the field theory. The resulting theories display gravity
descriptions of, amongst other properties, confinement, fermion
condensates, instantons, and thermal phase transitions. In this
paper we will address another familiar aspect of field theory,
Cooper pair formation in a fermionic system at high density.

A number of authors \cite{spin, thermodynamic}
have already studied the $AdS$/CFT correspondence at high
density though they have concentrated on the scalar sector of the
${\cal N}=4$ gauge theory. The system may be placed at high density by the
inclusion of a chemical potential. The chemical potential may be
thought of as the vev of the temporal component of a spurious
gauge field associated with a conserved global $U(1)$ symmetry. It
is natural to pick a $U(1)$ subgroup of the global $SU(4)_R$ group
of the ${\cal N}=4$ theory. In the gravity dual the $SU(4)_R$ symmetry
appears as a gauged symmetry of the IIB supergravity on $AdS_5
\times S^5$. As was recognized in \cite{spin} the gauge field
originates in the 10 dimensional supergravity theory as an element
of the metric and can be made non-zero by spinning the D3 branes
of the construction on the $S^5$ space. This can be naively seen by
starting with a static construction and performing a boost to a
slowly rotating reference frame where an angular coordinate in some
plane takes the form $\phi' = \phi + \omega t$. There are now $\phi' t$
components of the Lorentz transformation matrix $\Lambda_{ij}$
and hence $\phi' t$ components in the metric.

The phenomena we wish to study is Cooper pair formation. High
density fermionic systems have of course been much studied in
condensed matter \cite{Polsuper}
but the relativistic analogues have also recently
become an area of interest in QCD where a rich phase structure of
colour superconductors has been uncovered \cite{CSC}. In both the
non-relativistic \cite{Polsuper} and relativistic cases \cite{CSC,CSCRG}
there are fairly rigorous
arguments that {\it in the presence of a Fermi surface any
attractive interaction gives rise to Cooper pair formation}. We
will briefly review the renormalization group \cite{Polsuper, CSCRG}
justification for
this and the form of the condensate expected as a result of the
gauge interactions in the ${\cal N}=4$ theory. The preferred condensate is
a colour singlet so the theory is a superfluid rather than a
colour superconductor.

Capturing this phenomena in the ${\cal N}=4$ SYM theory is complicated by
the scalar sector of the theory which also transforms under the
$SU(4)_R$ symmetry. When the chemical potential is introduced as a
spurious gauge field vev, scalar operators are also introduced that
destabilize the moduli space of the theory. In the gravity dual
this is clear; the D3 branes experience no potential on the six
dimensional space transverse to their world volume (corresponding
to the existence of the moduli space in the field theory) so there
is no central force that can be used to sustain rotational motion
(the source for the gauge field vev). Formally one should cure
this problem by the inclusion of a positive scalar mass or, as
previous authors have considered, by including finite temperature
which indirectly generates such a mass.

We shall make do with finding a metastable vacuum in the scalar
sector to avoid such complications. The D3 brane positions in the
transverse space are described by the vevs of the adjoint scalar
fields in the field theory on their surfaces and they can
therefore be separated in a non-commutative fashion at the expense
of energy of the form $tr[\phi^\dagger, \phi]^2$. We will use this
central force to stabilize rotation of the D3 branes and provide
us with the metastable vacuum. The construction is essentially
that in (M)atrix theory of a spinning (fuzzy) D2 brane with surface D0
charge introduced by Harmark and Savvidy \cite{harmark}
and we will follow their
methodology.  Expanding branes non-commutatively induces couplings
to higher dimensional RR forms as was realized by Myers \cite{myers}. The
resulting configuration of expanded D3 branes may be thought of as
a D5 brane with surface D3 brane charge.  It would be nice to find
supergravity backgrounds when all the D3 branes are spinning but
in this paper we shall restrict ourselves to the case where only a
small, probing, number of the D3 are spinning so the background
geometry remains $AdS$. In the field theory this will imply that we
have put a chemical potential in for a restricted set of gauge degrees
of freedom.
Nevertheless this will be sufficient to see
the important physical effect of Cooper pair formation in this
subsector of the theory.

The D5 acts as a source for the RR 6-form or its dual description,
a 2-form. This whole story is of course very analogous to the
Polchinksi Strassler \cite{ps} analysis of the supergravity dual
of the ${\cal N}=1^*$ gauge theory. There, a mass that breaks
${\cal N}=4$ to ${\cal N}=1$ in the infra-red is introduced by
placing the D3 brane construction in an appropriate background
3-form field strength. The 3-form provides a radial potential for
the D3 branes with a minimum resulting from the interplay with the
energy of the non-commutative expansion of the D3s. They find a
fermion condensate as a sub-leading term in the 3-form solution.
For our case this is the leading part of the 3-form. We show,
following their methods, that the symmetries and RG scaling
dimension of this induced form match with the formation of a
Cooper pair condensate. In fact there is a triple scalar operator
with the same R-charge and scaling dimension as the condensate
which therefore mixes making the precise field theory
interpretation difficult although the consistency of the story is
appealing.

The D3/D5 construction we produce has explicit time dependence
since the D5 topples in a six dimensional space and hence it spins
down by loss of energy to RR-form and gravity waves. Harmark and
Savvidy \cite{harmark} studied these emissions extensively for the construction
in Minkowski space and the same phenomena is apparent in $AdS$.
Since these forms correspond to sources and vevs of field theory
operators, the coupling of the world volume scalars 
to them may be thought of as higher dimension operators. These
operators allow the metastable vacuum to decay to the true
run-away vacuum and hence the angular momentum (R-charge) of the
construction eventually resides on the asymptotic edge of moduli
space.

A detailed study of the thermodynamics of the $AdS$/CFT
correspondence at finite temperature and density was made in
\cite{thermodynamic}.\footnote{We are very grateful to Clifford
Johnson for long discussion on this phase structure from which
this work emerged and also for introducing us to spinning branes
and their connection to the chemical potential.} The absence of a
superfluid phase was noted in \cite{hawking} but the instability
of our construction suggests that the phenomena will never be seen
in the true vacuum of the ${\cal N}=4$ theory.

\section{The Field Theory}

${\cal N}=4$ SYM theory may be placed at high density
by the inclusion of a chemical
potential, $\mu$, for the gauginos
\begin{equation} \label{fermlag}
\Delta L = i \mu \lambda_i T_{ij} \gamma_0 \lambda_j.
\end{equation}
The chemical potential may be thought of as the vev of the
temporal component of a spurious gauge field associated with a
conserved global $U(1)$ symmetry, with generator $T_{ij}$.
In the $AdS$/CFT it is natural to
pick a $U(1)$ subgroup of the global $SU(4)_R$ group of the ${\cal N}=4$
theory and in this paper we shall restrict ourselves to the
fundamental representation generator $T_{ij} = diag(1,1,1,-3)$. In
the gravity dual the $SU(4)_R$ symmetry appears as a gauged
symmetry of the IIB supergravity on $AdS_5 \times S^5$. As
discussed in the introduction the gauge field originates in the 10
$d$ supergravity theory as an element of the metric and
can be made non-zero by spinning the D3 branes of the construction
on the $S^5$ space. 
%In terms of the 6 dimensional representation
%of SU(4)$_R$, relevant to the symmetries of the $S^5$,
%the generator we study corresponds to diag(1,-1,1,-1,1,-1) or,
%in an SO(6) representation to an equal rotation in each of three SO(2)
%sub-groups/planes
%of the six dimensional space transverse to the D3 branes. 

In terms of $SO(6)$, the isometry group of the $S^5$,
the generator we study corresponds to an equal rotation in each of 
three $SO(2)$
sub-groups/planes
of the six dimensional space transverse to the D3 branes. 

As a result
of this motion we
will create a D5 brane wrapped on a 2-sphere.
The choice of different rotations in the three planes
generates ellipsoidal configurations as discussed in \cite{harmark}
but for simplicity we will restrict to the most symmetric case.

Giving a vev to this temporal gauge field does not simply induce
the term in (\ref{fermlag}) but also a scalar term coming from the
(spurious) scalar covariant derivative
\begin{equation}
| D^\mu \phi |^2 \rightarrow \sum_{i=1}^6 g^2 \mu^2 | \phi_i |^2,
\end{equation}
which is a positive quadratic term in the lagrangian and hence a
negative mass term in the potential (here $g$ is the gauge coupling). 
Since the ${\cal N}=4$ theory has a
scalar moduli space this term serves to destabilize it and the
theory is unbounded! The gravity dual of this phenomena may be
seen by attempting to place a spinning D3 brane probe in the $AdS$
background. The $AdS_5 \times S^5$ background is
\begin{equation}\label{ads}
ds^2 = H^{-1/2} dx_{//}^2 + H^{1/2} (dr^2  + r^2 d\Omega_5^2),
\hspace{1cm}
C_4 = H^{-1} dx^0 \wedge dx^1 \wedge dx^2 \wedge dx^3,
\end{equation}
where $x_{//}$ are the 3+1 dimensions of the D3 world volume and
$r$ and $\Omega_5$ describe the six transverse dimensions.
$H$ is $L^4/r^4$ in $AdS$, with $L$ the  radius of $AdS$ (in general
$H$ can be a more complicated harmonic function
in which case the background describes the ${\cal N}=4$
theory at a generic point in the moduli space).
Note that under a dilatation in the four dimensional $x^\mu$ space
the radial coordinate $r$ transforms with mass dimension one which indicates
its role as the direction of renormalization group flow.

The spinning D3 probe experiences the metric through the abelian
Dirac-Born-Infeld action
\begin{equation}
S = - \tau_3 \int d^4 \xi e^{-\phi} \sqrt{\det \left[ G_{ab} + 2\pi
    \alpha' F_{ab}
\right]} + \mu_3 \int C_4,
\end{equation}
where $\tau_3= g_s^{-1}\mu_3$
%= (2 \pi)^{-3} \alpha'^{-2}%
are the D3 brane tension and charge respectively, 
and $G_{ab}$ is the pullback of the metric\footnote{We use $M,N =
  0,\dots,9$ for the 10$d$ indices, $a,b=0,\dots,3$ 
for the world volume coordinates of the D3 brane and $m,n=4,\dots,9$ for the
coordinates transverse to the brane.}
\begin{equation}
G^{ab}= G_{M N} {\partial x^M \over \partial \xi^a}
{\partial x^N \over \partial \xi^b}.
\end{equation}
Switching on only the scalar fields ($x^m = 2 \pi \alpha' X_m $) 
and allowing slow rotation
on $S^5$ we find
\begin{equation}
L_{\rm probe} = {\mu_3 \over 2} r^2 \dot{\Omega}^2_5.
\end{equation}
There is only a kinetic term so the position of the D3 brane
(scalar vev) runs away to infinity as
the field theory analysis predicted.

To stabilize the scalar sector we shall concentrate on a metastable vacuum
where we allow the adjoint scalar fields to take non-commuting values.
The scalar potential is of the form
\begin{equation}
V = \sum_{i=1}^6 tr \left[ \varphi_i^\dagger, \varphi_i \right]^2,
\end{equation}
and clearly contributes a quartic term when the vevs are
non-commuting which will stabilize the negative mass term from the
chemical potential. The resulting scalar vev will be of order
$\mu$. We perform an analysis of this vacuum in the next section
in the D3 brane world volume theory.

Putting aside the scalar sector for the moment we can discuss what
we would expect to happen in the fermionic sector were it in
isolation. The dynamics of the theory should lie close to the
Fermi surface so it is natural to look at an effective theory
close to the Fermi surface following the analysis of
\cite{Polsuper, CSCRG}. We
write the fermions' momenta as a piece on the Fermi surface plus a
small contribution perpendicular $\vec{p} = \vec{k} + \vec{l}$.
Now as we scale towards the Fermi surface such that $l \rightarrow
s l$ and $E \rightarrow s E$, with the scaling factor $s
< 1$, for the fermion kinetic term to be marginal
\begin{equation}
\int dt d^3x \bar{\lambda} \partial^\mu \gamma_\mu \lambda,
\end{equation}
we require $\lambda \rightarrow s^{-1/2} \lambda$. Under these scalings a
four fermion interaction
\begin{equation}
G \int dt d^3p_1  d^3p_2 d^3p_3  d^3p_4  \bar{\lambda} \lambda \bar{\lambda}
\lambda \delta(\vec{p_1} + \vec{p_2} - \vec{p_3}- \vec{p_4})
\end{equation}
is naively irrelevant (scaling as $s$) since the delta function is
usually independent of $\vec{l}$ which are small. Only for
scatterings between two fermions with equal and opposite momenta
do the factors of $\vec{k}$ cancel in the delta function leaving
it dependent on $\vec{l}$ and hence the  operator marginal. Close
to the Fermi surface there is therefore a vast simplification of
the theory to these scatterings with special kinematics. In fact
this has the effect of simplifying loop diagrams in the same way
as the large $N$ expansion in a theory with four fermion
interactions. The running of the marginal operator is just through
the ``bubble sum'' diagrams and results in a logarithmic running
with an IR pole (whose position is dependent on the initial
strength of the coupling) where Cooper pairing is expected to
occur from the resulting strong interaction. In fact the gap
equation truncation of the Schwinger Dyson equation becomes exact
and the theory and its condensate are exactly soluble. Obviously
this four fermion theory analysis is naive for a gauge theory but
it is indicative of the expected behaviour in the presence of an
effective four fermion vertex from gluon exchange.

The preferred condensate turns out \cite{CSC,CSCRG} to be between
two like helicity spinors or, since they have opposite momentum,
the anti-symmetric spin singlet state. Not surprisingly the
preferred colour state is simply that with the most attractive
interaction before the running. In the case of ${\cal N}=4$ SYM the most
attractive channel between adjoint gauginos is the symmetric
colour singlet. The flavour structure is then determined by Fermi
Dirac statistics to be a symmetric state. In other words we expect
the condensate to be an element of the $4 \times 4 = 10$
dimensional representation of $SU(4)_R$.

\section{The Spinning Fuzzy D5 Brane}

In \cite{harmark} the solution for a spinning D2 brane with
surface D0 brane charge in flat spacetime was presented. In fact
the construction is equally valid for D3 branes non-commutatively
blowing up into a spinning D5, so we will review the construction
in \cite{harmark} applied to the D3 brane case which is relevant
to our problem. We will assume the D3 branes are flat in the
directions 0123 and study their motion in the six transverse
directions. The effective action for $N$ coincident D3 branes 
%may be obtained
%by dimensionally reducing ${\cal N}=1$ SYM in 10d or from the
%non-abelian Born Infeld action 
is given by \cite{myers}
\beq
S_{BI} = - \tau_3 \int d^4x Tr \left( e^{-\phi} \sqrt{
- det(P[E_{ab} + E_{ai} (Q^{-1} - \delta)^{ij} E_{jb}] + \lambda F_{ab})
det(Q^i_j)} \right),
\eeq

\noindent where $P$ indicates the pull back, $E_{ab} = G_{ab} +
B_{ab}$, $\lambda = 2\pi \alpha'$, and
\beq
Q^i_j = \delta^i_j  + i \lambda [ X^i, X^k] E_{kj}.
\eeq

\noindent The second term in the first determinant is zero when the
metric is diagonal.

The resulting action for the scalar fields in Minkowski space,
describing motion
in the 6 transverse dimensions ($i=4,...,9$), is given by
%\beq S = T \int d^4x  Tr \left( {1 \over 2} g_{tt} g_{ij}
%\dot{X}^i \dot{X}^j + {1 \over 4} {1 \over (2 \pi l_s^2)^2} g_{ik}
%g_{jl} [ X^i, X^j][ X^k, X^l]   \right) \eeq
%
\beq S = \tau_3 \int d^4x  Tr \left( {1 \over 2}
\dot{X}^i \dot{X}^i + {\lambda \over 4}
[ X^i, X^j][ X^i, X^j]   \right). \eeq

\noindent   The essence of the construction is to use the positive potential
from the non-commutativity to provide the centrifugal force to
sustain rotational motion. It is not possible to support a
2-sphere by rotation within the world volume of the 2-sphere.
Instead we must allow the 2-sphere to topple in the 6$d$ space. We embed  the 
2-sphere in the directions 468. Then we pair the
three axes each with an axis in the transverse 3$d$ space 579 and
allow rotation in each of the three resulting planes. The ansatz
for the $N$ D3 to have non-commutatively puffed up and to be
spinning in the three separate transverse planes (45, 67, 89) with
the same angular velocity $\omega$ is \cite{harmark}
\beq
\begin{array}{ccc}
X_4(t) = {2 \over \sqrt{N^2 - 1}} T^1 r_4(t)&, & X_5(t) = {2
\over \sqrt{N^2 - 1}} T^1 r_5(t), \\ X_6(t) = {2 \over \sqrt{N^2 -
1}} T^2 r_6(t)&, & X_7(t) = {2 \over \sqrt{N^2 - 1}} T^2 r_7(t),
\\ X_8(t) = {2 \over \sqrt{N^2 - 1}} T^3 r_8(t)&, & X_9(t) = {2
\over \sqrt{N^2 - 1}} T^3 r_9(t),
\end{array}
\eeq 
where the $T^i$ are the three generators of
$SU(2)$ in the $N\times N$ irreducible representation. 
They have the properties
\beq [ T^i, T^j] = i \epsilon_{ijk} T_k, \hspace{1cm} \sum_i T_i^2
= {N^2-1 \over 4} {\bf I}, \hspace{1cm} Tr(T_i^2) = {N(N^2-1)
\over 12}. \eeq

Substituting into the action,  using a Minkowski metric, produces a
lagrangian
\beq L = {N \tau_3 \over 3} \left( {1 \over 2} \sum_{i=4}^9 \dot{r}_i^2
- {\alpha^2 \over 2 } \left[ (r_4^2 + r_5^2)(r_6^2 + r_7^2) +
(r_4^2 + r_5^2)(r_8^2 + r_9^2) + (r_6^2 + r_7^2)(r_8^2 +
r_9^2)\right] \right), 
\eeq
where $\alpha = 1/( \pi l_s^2\sqrt{N^2-1}$). The equations of
motion are thence
\beq \begin{array}{c} \ddot{r}_{4(5)} = - \alpha^2 (r_6^2 + r_7^2
+ r_8^2 + r_9^2 )r_{4(5)} \\ \ddot{r}_{6(7)} = - \alpha^2  (r_4^2
+ r_5^2 + r_8^2 + r_9^2 )r_{6(7)}
\\ \ddot{r}_{8(9)} = - \alpha^2  (r_6^2 + r_7^2 +
r_4^2 + r_5^2 )r_{8(9)},
\end{array}
\eeq

\noindent which have solutions
\beq \begin{array}{c} r_4 = R \cos \omega t,  \hspace{1cm} r_5 = R
\sin \omega t, \\ r_6 = R \cos \omega t,  \hspace{1cm} r_7 = R \sin
\omega t, \\ r_8 = R \cos \omega t,  \hspace{1cm} r_9 = R \sin
\omega t, \\
\end{array}
\eeq
\noindent with
\beq \omega = \sqrt{2} \alpha R. \eeq

We have then a stable solution
of the form we seek. Ideally we should now find the background
geometry in the large $N$ limit and take the near horizon limit to
determine the dual field theory behaviour. However, this is
clearly a huge task since the construction has time dependent
motion and couples to the metric, the RR 6-form and RR 4-form
potentials. It is not even clear a priori that such a solution
would exist as a  self consistent solution although it is likely.
Instead we will retreat to an easier problem that will
nevertheless display the important physics. Instead of spinning
all $ N $ D3 branes we will choose instead to spin $n \ll N$
corresponding to introducing a chemical potential for only $n$
of the gauge degrees of freedom. In terms of the D3 computation we
will therefore be able to work in an $n/N$ expansion. Essentially
we are including a probe D5 brane with surface D3 charge. To
leading order the background metric is that of $AdS_5 \times S^5$ in
(\ref{ads}). Inserting the background into the non-abelian Born Infeld action
we find
\beq
S_{BI} = - \tau_3 \int d^4x  Tr  \sqrt{H^{-2} (1 + H \dot{X}_i)} (
1  - {\lambda^2 \over 4} H [X^i, X^j]^2) + \tau_3 \int d^4x H^{-1} .
\eeq

The potential term vanishes as usual and in the remaining leading terms
the factors of $H$ cancel leaving precisely the Minkowski space action
we had before. This cancellation is to be expected since the action should
describe the ${\cal N}=4$ Yang Mills theory where $H$ is absent.
The computation in $AdS$, therefore, exactly mirrors \cite{harmark}
and the construction is seen to
also exists in $AdS$. We
will now go on to analyze the effects of the construction on the
background supergravity solution to see if a fermion condensate is
indeed present.

\section{Asymptotic Operators}

Having determined that there is  a meta-stable vacuum where the D3
branes balance their rotational motion against non-commutative
expansion, we wish to ask what operators in addition to the
chemical potential characterize the vacuum. 
There is a D5 source in the interior of
the space which will give rise to a non-zero 3-form.
To see this we look at the 
supergravity equations of motion linearized around the $AdS_5 \times S^5$
background.
At linear order in the $n/N$ expansion we can neglect the 
back-reaction of the brane on the metric and the RR 4-form, and 
simply consider the linearized equation for a 3-form field
\beq
G_3=F_3 -\hat{\tau} H_3.
\eeq

Here $F_3$ ,$H_3$ are the RR and NS 3-form, and $\hat{\tau}=i/g$ 
is the background value of the dilaton field.
%In an expansion in
%$n/N$ the asymptotic metric at  leading order will be that of
%$AdS_5 \times S^5$. 
%There is though a D5 source in the interior of
%the space which will give rise to a non-zero 3-form.
The linearized equation of motion and Bianchi identity for the 
field $G_3$ were shown by
Polchinksi and Strassler \cite{ps} to take the simple form\footnote{In
  the following we set the $AdS$ radius, $L$, to one.}
\begin{eqnarray}
d(r^4 (*_6 G_3 - i G_3) ) = 0,\nonumber\\
d G_3 = J_4, 
\end{eqnarray}
where $*_6$ indicates dualizing in the the six dimensional
transverse space using a flat metric for contractions, and $J_4$ is 
the D5 brane source.

Consider first a static D5 brane lying in $R^4$, and wrapped as a
2-sphere in the transverse six dimensional space \cite{ps}. We call the
three coordinates in which the 2-sphere lies the $w$ directions
and it is then at the origin in the remaining three $y$
directions. It acts as an electric source for the 6-form potential
with components in the $0...4, \theta_w, \phi_w$ directions.
Dualizing to find the magnetic source for the 3-form gives
\begin{equation}
J_4 =  4 \pi^2 \alpha' \delta^3(y) \delta(w-r_0) d^3y
\wedge dw
\end{equation}
where $w$ is the radial coordinate on the sphere and $r_0$ is the radius
of the sphere which we saw above is proportional to the angular velocity.

It is convenient to write $G_3$ in terms of a potential
\begin{equation}
G_3 = *_6 d w_2 + i d w_2.
\end{equation}
Working in the gauge $d w_2 =0$, this ansatz solves the
equation of motion leaving just the Bianchi identity with source
\begin{equation} \label{gauss}
\partial_m \partial_m w_2 = {2 \pi^2 \alpha^\prime  \over r_0} \delta^3(y)
\delta(w-r_0) \epsilon_{ijk} w^i dw^j \wedge dw^k,
\end{equation}
which has the asymptotic solution \cite{ps}
\begin{equation}
w_2 \sim - {8 \alpha^\prime  r_0^3 \over 3 r^6} \epsilon_{ijk} w^i dw^j
\wedge dw^k.
\end{equation}
The solution scales (in an inertial frame transverse to the radial
direction) as $1/r^3$, the normalizable solution for 
 an operator of dimension 3. This is the appropriate behaviour for
the gaugino Cooper pair condensate we seek.

We must also check that the symmetry properties are correct. Again following
Polchinski Strassler \cite{ps} we adopt complex coordinates
\begin{equation}
z_1 = { w^1 + i y^1 \over \sqrt{2}}, \hspace{1cm} z^2 = { w^2 + i
y^2 \over \sqrt{2}}, \hspace{1cm} z_3 = { w^3 + i y^3 \over
\sqrt{2}}.
\end{equation}
Under a rotation $z^i \rightarrow e^{i \phi_i} z_i$ the gauginos transform as
\begin{equation}
\begin{array}{c}
\lambda_1 \rightarrow e^{i ( \phi_1 - \phi_2 - \phi_3)/2} \lambda_1, \\
\lambda_2 \rightarrow e^{i ( -\phi_1 + \phi_2 - \phi_3)/2} \lambda_2, \\
\lambda_3 \rightarrow e^{i ( -\phi_1 - \phi_2 + \phi_3)/2} \lambda_3, \\
\lambda_4 \rightarrow e^{i ( \phi_1 + \phi_2 + \phi_3)/2} \lambda_4.
\end{array}
\end{equation}

Thus we can construct a 3-form with the same symmetry
transformations as a condensate
\begin{equation}
\langle \lambda_1 \lambda_1 \rangle dz^1 \wedge d\bar{z}^2 \wedge d\bar{z}^3
+ \langle \lambda_2 \lambda_2 \rangle d\bar{z}^1 \wedge dz^2 \wedge d\bar{z}^3
+ \langle \lambda_3 \lambda_3 \rangle d\bar{z}^1 \wedge d\bar{z}^2 \wedge dz^3
+ \langle \lambda_4 \lambda_4 \rangle dz^1 \wedge dz^2 \wedge dz^3.
\end{equation}
When all four condensates are equal the 3-form, written in the
real coordinates, becomes
\begin{equation}
 \langle \lambda \lambda \rangle (dw^1 \wedge dw^2 \wedge dw^3
+ i  dy^1 \wedge dy^2 \wedge dy^3).
\end{equation}

We can now compare this to the solution we have for the 3-form
field strength. The leading term has an epsilon tensor in the
three $w$ directions and hence has the correct symmetry properties
to correspond to a real condensate of this form. This is
essentially the result we were looking for - the introduction of a
chemical potential, in a vacuum where the scalar instability has
been removed, results in fermionic Cooper pair formation. The
precise form of the condensate is though somewhat surprising. We
introduced a chemical potential for the fermions of the form
$diag(1,1,1,-3)$ and hence would have expected only an $SU(3)$
symmetry in the condensate rather than the $SU(4)$ symmetry
observed. We note that the same lifting of the symmetry was
observed in the ${\cal N}=1^*$ theory \cite{ps}-
there the $SU(4)_R$ symmetry was
broken by a mass term for three of the four fermions. As noted there the
interpretation is clouded by a three scalar operator also in the
10 representation ($6 \times 6 \times 6 = 10 + ...$) of $SU(4)_R$
which can mix with the fermion condensate. In fact it makes no sense to
distinguish these operators in the theory but it is nevertheless
encouraging that such an operator is present.

In this analysis we have neglected the spin of the D5 which, however, 
is easily included. Equation (\ref{gauss}) becomes the wave equation
\begin{equation}
(\partial_m^2 - \partial_t^2) w_2 =
{2 \pi^2 \alpha^\prime  \over r_0} \delta^3(Im(ze^{i\omega t}))
\delta(Re(ze^{i\omega t})-r_0) \epsilon_{ijk} z^i dz^j \wedge dz^k
e^{i \omega t},
\end{equation}
where $z$ are the complex coordinates. The asymptotic solution is
\beq \label{besselsol}
w_2 \sim  - {8 \alpha'  r_0^3 \over  3 }z^i  {J_3( \omega r)
\over r^3}e^{i \omega t},
\eeq
where $J_3$ is one of the Bessel functions of the second kind (we
choose this solution since this Bessel function diverges in the interior).
Expanding the Bessel function for  small $\omega$ gives
\beq
w_2 \sim  - {8 \alpha'  r_0^3 \over  3  } z^i {F( \omega r)
\over r^6}  e^{i \omega t},
\eeq
where $F(\omega r) = 1+ {\cal O}(\omega^2 r^2)$ is
an oscillatory function in $r$.

The result still describes a condensate except that there is a
sinusoidal oscillation between the $w$ and $y$ directions. In
terms of the condensate this implies a time dependent shift in the global
phase. The spinning configuration is in fact giving off waves of
$C_2$ potential precisely of the type found in
\cite{harmark}
in Minkowski space. The angular momentum ($U(1)_R$ charge in the field theory)
is radiating away from the configuration and out through the $AdS$
space. A full analysis would also reveal $C_4$ and
gravitational radiation. Over time this radiation will spin the
construction down to a static state with the angular momentum lost
to the RR potentials radiating away. What is the interpretation of
this phenomena in terms of the dual field theory?

We know of course that the construction is only a metastable
vacuum with the true vacuum being a runaway in the scalar vev. We
must therefore expect to see the construction decay. The decay can
occur either through tunnelling or via higher dimension couplings
between the scalars in the D5 brane's world volume and other
unbounded operators in the theory. These operators are precisely
what the couplings to the RR forms represent. The RR form vevs are
sources describing the full set of primary operators in the theory
so the radiation indeed describes the decay of the metastable
vacuum to runaway operators. As in the case where the angular
momentum is endowed to commutative D3 branes the signal of the
runaway is that the angular momentum is carried to the edge of the
moduli space - in that case by the D3 brane motion whilst in this
more complicated case by the RR forms.

\section{Discussion}

We have shown that a fuzzy D5 sphere may be supported in the
$AdS$/CFT correspondence by rotation  on the $S^5$ sphere of a
small number of the D3 branes. In the field theory the rotation
corresponds to the inclusion of a chemical potential putting the
theory at high density with respect to a $U(1)_R$ symmetry group.
The ${\cal N}=4$ theory is unbounded in the scalar sector by a
chemical potential but this state represents a metastable vacuum
where the negative mass is balanced against a positive potential
from non-commutativity of the scalar fields. The D5 brane
necessarily couples to a $C_6$ (or dual $C_2$) potential the form
of which we have shown corresponds to an operator in the field
theory with the dimension and symmetry properties of a fermion
condensate. This matches our expectation that at high density
Cooper pair formation should result from the attractive gauge
interactions. In the future it remains as a challenge  to
establish the existence of such a configuration when all the D3
branes are rotating and find the full supergravity background. It
would also be nice to study the phenomena in a theory, with a
gravity dual, where there are no scalar fields (such as ${\cal
N}=1$ Yang Mills theory) since in the ${\cal N}=4$ theory scalar
operators mix with the fermion condensate muddying the
interpretation.

The rotating D5 brane is in fact a time dependent structure since
it topples in the $S^5$ space. This results, through its couplings
to RR-forms, in its decay by the emission of waves. We have
interpreted this as decay of the metastable vacuum to the true
unbounded vacuum of the ${\cal N}=4$ theory at high density,
though it would be interesting to further study the operators
present at asymptotic $r$ in (\ref{besselsol}) which presumably
describe the operators of the runaway vacuum.
%Note that such operators
%and decay will be present for any time dependent construction in $AdS$
%and it would be particularly interesting to study the giant graviton
%\cite{giantgrav} in this
%respect.

\vskip .2in
\noindent
{\bf Acknowledgements}\vskip .1in
\noindent
The authors are very much indebted to Clifford Johnson for lengthy discussions
and exploration of many dead ends off the final path of this paper. We 
also thank Alberto Zaffaroni for comments on the manuscript.
N.E is grateful for the support of a PPARC Advanced Fellowship.

\end{document}